\documentclass{amsart}
\usepackage{amssymb}

 \makeatletter

 \newcommand\setof[1]{\mathopen\{\,#1\,\mathclose\}}
 \newcommand {\sbs}{\subset}
 \newcommand{\half}{\frac{1}{2}}

 \newcommand \ag{\alpha}
 \newcommand \bg{\beta}
 \newcommand \dg{\delta}
 \newcommand \eps{\varepsilon}
 \newcommand \kg{\kappa}
 \renewcommand \lg{\lambda}
 \newcommand \sg{\sigma}

 \newcommand \Og{\Omega}

 \newcommand{\cH}{\mathcal{H}}
 \newcommand{\cQ}{\mathcal{Q}}
 \newcommand{\cS}{\mathcal{S}}

 \newcommand{\ZZ}{\mathbb{Z}}

 \newcommand\ang[1]{{\langle #1\rangle}}
 \newcommand\bra[1]{{\langle #1|}}
 \newcommand\ket[1]{{| #1 \rangle}}
 \newcommand\braket[2]{{\langle #1 | #2 \rangle}}

 \newcommand{\df}[1]{{\rmfamily\itshape\mdseries#1}}

 \let\ol\overline
 \let\ul\underline
 \let\prl\parallel
 \newcommand\nm[1]{{\| #1\|}} 

\def\subsection{\@startsection{subsection}{2}%
  \z@{.5\linespacing\@plus.7\linespacing}{-.5em}%
  {\normalfont\sffamily\bfseries\upshape}}

 \newcommand{\customqed}[1]{{\renewcommand{\qedsymbol}{#1}\qed}}
 \newcommand{\varqed}{\customqed{\hbox{$\diamondsuit$}}}

 \numberwithin{equation}{section} 

\newtheoremstyle{sfdefinition}
  {5pt}
  {5pt}
  {}
  {}
  {\sffamily\mdseries\upshape}
  {.}
  {.5em}
  {}

\newtheoremstyle{sftheorem}
  {5pt}
  {5pt}
  {\itshape}
  {}
  {\sffamily\mdseries\upshape}
  {.}
  {.5em}
  {}

 \theoremstyle{sftheorem}

 \newtheorem{lemma}{Lemma}[section]
 \newtheorem{theorem}{Theorem}
 \newtheorem{proposition}[lemma]{Proposition}

 \theoremstyle{remark}

 \newtheorem{Example}[lemma]{Example}
 \newtheorem{Examples}[lemma]{Examples}
 
 \newenvironment{example}{%
   \begin{Example}}{\varqed\end{Example}}

 \newtheorem{Remark}[lemma]{Remark}
 \newtheorem{Remarks}[lemma]{Remarks}
 
 \newenvironment{remark}{%
   \begin{Remark}}{\varqed\end{Remark}}
 \newenvironment{remarks}{%
   \begin{Remarks}}{\varqed\end{Remarks}}

 \newcommand\Prob{\mathop{\operator@font Prob}}
\def\paragraph{\@startsection{paragraph}{4}%
  \z@{.5\linespacing\@plus.7\linespacing}{-.5em}%
  {\normalfont\itshape}}

  \renewcommand{\le}{\leqslant}
  \renewcommand{\ge}{\geqslant}

 \newcommand{\lea}{\stackrel{+}{<}}
 \newcommand{\gea}{\stackrel{+}{>}}
 \newcommand{\eqa}{\stackrel{+}{=}}
 
 \newcommand{\lem}{\stackrel{\ast}{<}}
 \newcommand{\gem}{\stackrel{\ast}{>}}
 \newcommand{\eqm}{\stackrel{\ast}{=}}

 \newcommand{\tr}{\operatorname{Tr}}
 \let\dgr\dagger

 \newcommand{\bmu}{\boldsymbol{\mu}}
 \newcommand{\bkg}{\boldsymbol{\kappa}}
 
 \newcommand{\bid}{\mathbf{1}}
 \newcommand{\m}{\mathbf{m}}

 \newcommand{\Kq}{\text{\rmfamily\mdseries\upshape Kq}}
 \newcommand{\QC}{\text{\rmfamily\mdseries\upshape QC}}
 \newcommand{\Hu}{\overline H}
 \newcommand{\Hl}{\underline H}

 \makeatother

\begin{document}

 \title{Quantum algorithmic entropy}

  \author{Peter G\'acs}\thanks{The paper was written during the author's
visit at CWI, Amsterdam, partially supported by a grant of NWO}
  \address{CWI and Boston University}
  \email{gacs@bu.edu}
  \date{\today}

 \begin{abstract}
We extend algorithmic information theory to quantum mechanics,
taking a
universal semicomputable density matrix (``universal probability'')
as a starting point, and define complexity (an operator)
as its negative logarithm.

A number of properties of Kolmogorov complexity extend naturally 
to the new domain.
Approximately, a
quantum state is simple if it is within a small distance from a
low-dimensional subspace of low Kolmogorov complexity.
The 
von Neumann entropy of a computable density matrix is within an additive
constant from the average complexity.
Some of the theory of randomness translates to the new domain.

We explore the relations of the new quantity to the quantum Kolmogorov
complexity defined by Vit\'anyi (we show that the latter is sometimes as
large as $2n - 2\log n$) and the qubit complexity defined by
Berthiaume, Dam and Laplante.
The ``cloning'' properties of our complexity measure are similar to those
of qubit complexity.
 \end{abstract}


\maketitle

\section{Introduction}\label{s.intr}

Kolmogorov complexity (or by a more neutral name, description complexity)
is an attractive concept, helping to shed light onto such subtle concepts
as information content, randomness and inductive inference.
Quantum information theory, a subject with its own conceptual difficulties,
is attracting currently more attention than ever before, due to the
excitement around quantum computing, quantum cryptography, and the many
connections between these areas.
The new interest is also spurring efforts
 to extend the theory of description complexity
to the quantum setting: see~\cite{VitanyiThreeAppr00},
\cite{BerthDamLaplante00}. 
We continue these efforts in the hope that the correct notions 
will be
found at the convergence of approaches from different directions.
This has been the case for
the theory of classical description complexity and randomness, 
What we expect from these researches is an eventual deeper
understanding of quantum information theory itself.

One of the starting points from wich it is possible to arrive at description
complexity is Levin's concept of a universal semicomputable (semi)measure.
We follow this approach in the quantum setting, where probability measures
are generalized into density matrices.

In contrast to the
works~\cite{VitanyiThreeAppr00}, \cite{BerthDamLaplante00} we do not find
the notion of a quantum computer essential for this theory, even
to the notions and results found in these works.
The reason is that limitations on computing time do not play a role
in the main theory of description complexity, and given enough time, a
quantum computer can be simulated by a classical computer to any desired
degree of precision.

\subsection{Notation}

It seems that universal probability can also be defined in an
infinite-dimensional space (it should be simple to extend the
notions to Fock space), but we will confine ourselves to
finite-dimensional spaces, in order to avoid
issues of convergence and spectral representation for
infinite-dimensional operators.
Let us fix for each $N$ a finite-dimensional Hilbert space $\cH_{N}$,
with a \df{canonical orthonormal basis}
$\ket{\bg_{1}},\dots,\ket{\bg_{N}}$. 
(We do not use double index here, since we can assume that
$\cH_{N} \sbs \cH_{N+1}$ and the canonical basis of $\cH_{N}$ is also the
beginning of that of $\cH_{N+1}$.)
Let $\cQ_{n} = \bigotimes_{i=1}^{n} \cQ_{1}$ be the Hilbert space of $n$
qubits.
Let $\ket{0},\ket{1}$ be some fixed orthonormal
basis of $\cQ_{1}$.
Let $\ZZ_{2}^{n}$ be the set of binary sequences of length $n$.
If $x \in \ZZ_{2}^{n}$ then $x = (x(1),x(2),\dots, x(n))$, and we write
 \[
   l(x) = n.
 \]
We denote, as usual, for $x \in \ZZ_{2}^{n}$:
 \[
   \ket{x} = \bigotimes_{i=1}^{n} \ket{x(i)}.
 \]
We identify $\cQ_{n}$ with $\cH_{2^{n}}$, with the canonical basis element
$\ket{\bg_{x}}=\ket{x}$.

If we write $\psi$ or $\ket{\psi}$
for a state then the corresponding element of the
dual space will be written either as $\psi^{\dgr}$ or as $\bra{\psi}$.
Accordingly, the inner product can be written in three ways as
 \[
   \braket{\phi}{\psi} = \ang{\phi, \psi} = \phi^{\dgr} \psi.
 \]
As usual, we will sometimes write 
 \[
\ket{\phi}\otimes\ket{\psi} = \ket{\phi}\ket{\psi} = \ket{\phi,\psi}.
 \]
The operation $\tr$ denotes trace, and over a tensor product space
$\cH_{X} \otimes \cH_{Y}$, the operation $\tr_{Y}$ denotes partial trace.

As usual, for self-adjoint operators $\rho, \sg$, let us write
$\rho \le \sg$ if $\sg - \rho$ is nonnegative definite.

Let us call a quantum state $\ket{\psi}$, with coefficients 
$\braket{\bg_{i}}{\psi}$ that are
\emph{algebraic numbers}, \df{elementary}.
The reason for going to coefficients that are algebraic numbers is that
this allows us the usual operations of linear algebra (orthogonalization,
finding eigenvalues and eigenvectors) while remaining in the realm of
elementary objects.

Whenever we write $U(p) = \ket{\phi}$ for a Turing machine $U$, we mean that
$U$ simply outputs the (algebraic definitions of the)
coefficients of the elementary state $\ket{\phi}$.
Similarly, let us call a self-adjoint operator
$T$ \df{elementary} if it is given by a matrix with algebraic entries.

We will also write $U(p) = \ket{\phi}$ if $U(p)$ outputs a sequence
of tuples $(c_{1k}, \dots, c_{Nk})$ for $k=1, 2, \dotsc$, where
$c_{ik}$ is an elementary approximation of $\braket{\bg_{i}}{\phi}$ to
within $2^{-k}$.
In this case, we say that $\ket{\phi}$ is a \df{computable} quantum state with
program $p$.
We can talk similarly about a program computing a linear operator on the
finite-dimensional space, or even computing an infinite sequence
 $\ket{\phi_{1}},\ket{\phi_{2}}, \dotsc$ of states, in which case we output
progressively better approximations to more and more elements of the
sequence.

Let $\lea$ denote inequality to within an additive constant, and 
$\lem$ inequality to within a multiplicative constant.

We assume that the reader knows the definition and simple properties 
of Kolmogorov complexity,
even the definition of its prefix-free version $K(x)$.
For a reference, use~\cite{LiViBook93}.

\subsection{Attempts to define a quantum Kolmogorov complexity}

In~\cite{VitanyiThreeAppr00}, a notion of the description complexity of a
quantum state was introduced.
Though that definition uses quantum Turing machines, this does not seem
essential.
Indeed, a quantum Turing machine can simulate a classical one.
And if there is no restriction on computing time then any state 
output by a quantum Turing machine starting from $\ket{0\dots 0}$ can
also be output with arbitrary approximation by some ordinary Turing
machine.
We reproduce the definition from~\cite{VitanyiThreeAppr00} as follows.
For $\ket{\psi} \in \cH_{n}$, let
 \[
 \Kq(\ket{\psi} \mid N) 
   = \min \setof{l(p) - \log |\braket{\phi}{\psi}|^{2} : U(p, N) = \ket{\phi}}.
 \]
So, the complexity of $\ket{\psi}$ is made up of the length of a program
describing an approximation $\ket{\phi}$ to $\ket{\psi}$ and a term penalizing for bad
approximation.
It is proved in~\cite{VitanyiThreeAppr00} that for $\ket{\psi} \in \cQ_{n}$,
 \[
   \Kq(\ket{\psi} \mid n) \lea 2n.
 \]
The lower bounds given in that paper are close to $n$.
The following theorem will be proved in Section~\ref{s.Kq-proof}.

 \begin{theorem}\label{t.Kq-lb}
For large enough $n$, there are states
$\ket{\psi} \in \cQ_{n}$ with $\Kq(\ket{\psi} \mid n) > 2 n - 2\log n$.
 \end{theorem}

An entirely different approach to quantum Kolmogorov complexity is
used in~\cite{BerthDamLaplante00}, where even the defining programs
consist of qubits rather than ordinary bits.
I will refer informally to 
complexity defined in~\cite{BerthDamLaplante00} as ``qubit complexity''.
Despite the difference in some of the goals and basic definitions,
still a number of results of that paper look somewhat similar to ours.

\subsection{This paper}

The definition of $\Kq$ reflects the view that quantum states
should not be accorded the status of individual outcomes of experiments,
and therefore $\Kq$ strives only to approximate specification.
We go a little further, and approach quantum complexity using
probability distributions to start with.
We find a
universal semicomputable (semi-) density matrix (``universal probability'')
and define a ``complexity operator'' as its negative logarithm.
Depending on the order of taking the logarithm and the expectation, two
possible complexities are introduced for a quantum state $\ket{\psi}$:
$\Hl(\ket{\psi}) \lea \Hu(\ket{\psi})$.

A number of properties of Kolmogorov complexity extend naturally 
to the new domain.
Approximately, a
quantum state is simple if it is within a small distance from a
low-dimensional subspace of low Kolmogorov complexity.
(Ideally, the three vague terms
should play a role in the following decreasing
order of significance: dimension, complexity, closeness.)
This property can be used to relate our algorithmic entropy to both
Vit\'anyi's complexity and qubit complexity.
We find that $\Hl$ is within constant factor of Vit\'anyi's
complexity, that
$\Hu$ essentially lowerbounds qubit complexity and upperbounds an oracle
version of qubit complexity.

Though Vit\'anyi's complexity is typically 
close to $2n$, while qubit complexity is $\lea n$, these are
differences only within a constant factor; on the other
hand, occasionally $\Hl$ can be much smaller than $\Hu$ and thus
Vit\'anyi's complexity is occasionally much smaller than qubit complexity.
This is due to the permissive way in which 
Vit\'anyi's complexity deals with approximations.

The 
von Neumann entropy of a computable density matrix is within an additive
constant from the average complexity.
Some of the theory of randomness translates to the new domain, but new
questions arise due to non-commutativity.

The results on the maximal complexity of clones are sharp, and
similar to those in~\cite{BerthDamLaplante00}.

\section{Universal probability}

Let us call a nonnegative 
real function $f(x)$ defined on strings a \df{semimeasure} if 
$\sum_{x} f(x) \le 1$, and a \df{measure} (a probability distribution)
if the sum is 1.
A function is called \df{lower semicomputable} if there is a 
monotonically increasing sequence $g_{n}(x)$ of functions converging to it
such that $(n, x) \mapsto g_{n}(x)$ is a computable function mapping into
rational numbers.
It is computable when it is both lower and upper semicomputable.
(A lower semicomputable measure can be shown to be also computable.)
The reason for introducing semicomputable semimeasures is not that
computable measures are not felt general enough; rather, this step is
analogous to the introduction of recursively enumerable sets and partial
recursive functions.
Just as there are ``universal'' (or, ``complete'' in terms of, say,
many-one reduction) recursively enumerable sets but no universal recursive
sets, there is a universal semicomputable semimeasure in the sense of the
following proposition, even though there is no universal computable
measure.

Let $U$ be an optimal prefix Turing machine used in the definition of
$K(x)$, and let $z_{1},z_{2},\dotsc$ be an infinite sequence.
Then the quantity $U(z)$ is well-defined: it is the output of $U$
when $z$ is written on the input tape.
Let $Z_{1},Z_{2},\dotsc$ be an infinite coin-tossing 0-1 sequence, and let
us define 
 \begin{equation}\label{e.random-program}
  \m'(x) = \Prob[U(Z) = x].
 \end{equation}

 \begin{proposition}[Levin]\label{p.univ}
There is a semicomputable semimeasure $\mu$ with the property that for
any other semicomputable semimeasure $\nu$ there is a constant $c_{\nu} > 0$
such that for all $x$ we have $c_{\nu} \nu(x) \le \mu(x)$.
Moreover, $\mu \eqm \m'$.
 \end{proposition}
 \begin{proof}[Proof sketch]
We define a Turing machine $T$ that will output a sequence 
$(p_{t}, x_{t}, r_{t})$ where $r_{t}$ is a positive rational number.
At any time $t$, let $r_{t}(p,x)$ be defined as follows.
If there is no $i\le t$ for which some $(p,x, r_{i})$ has been outputted then
$r_{t}(p,x)=0$; otherwise, $r_{t}(p,x)$ is the maximum of those $r_{i}$.
The machine $T$ will have the following property for all $p$:
 \begin{equation}\label{e.T-sum-cond}
   \sum_{x}r_{t}(p,x) \le 1.
  \end{equation}
To define $T$, take a universal Turing machine $V(p,x,n)$.
Let $T$ simulate $V$ simultaneously on all inputs.
If at any stage of the simulation, some $V(p,x,n)$ has been found,
then $T$ checks whether it can interpret $V(p,x,n)$ as a positive rational
number $r$, and whether it can output the triple $(p,x,r)$ while keeping
the condition~\eqref{e.T-sum-cond}.
If yes, the triple is outputted, otherwise it is not, and the simulation
continues.
Define $\nu(p,x) = \lim_{t} r_{t}(p,x)$.
Then it is easy to check that $\mu(x) = \sum_{p} 2^{-p-1} \nu(p,x)$ satisfies
the conditions of the proposition.

To show $\mu \eqm \m'$, note that 
the random variable whose distribution is $\mu$
can be represented as a function of the
coin-tossing infinite sequence.
It is not difficult to check that the
function in question now can be implemented by a prefix Turing machine.
 \end{proof}

We will call any semicomputable semimeasure $\mu$ with the property in the
proposition ``universal''.
Any two universal semimeasures dominate each other within a multiplicative
constant.
We fix one such measure and denote it by 
 \[
  \m(x)
 \]
and call it the \df{universal probability}.
Its significance for complexity theory can be estimated by
by the following theorem, deriving the prefix complexity $K(x)$
from the universal probability.

\begin{proposition}[Levin's Coding Theorem] We have $K(x) = -\log \m(x)$.
\end{proposition}

The lower bound $(-\log \m(x)) \lea K(x)$ comes easily from the fact
that $K(x)$ is upper semicomputable and satisfies the ``Kraft inequality''
$\sum_{x} 2^{-K(x)} \le 1$.
For the proof of the upper bound, see~\cite{LiViBook93}.

The above concepts and results can be generalized to the case
when we have an extra parameter in the condition:
we will therefore talk about $\m(x \mid N)$, the universal probability
conditional to $N$, a function maximal within a
multiplicative constant among all lower semicomputable
functions $f(x, N)$ which also satisfy the condition
$\sum_{x} f(x, N) \le 1$.
The coding theorem generalizes to $2^{-K(x \mid N)} \eqm \m(x \mid N)$.

Constructive objects other than integers or strings
can be encoded into integers in some canonical way.
Elementary
quantum states $\ket{\psi} \in \cH_{N}$ also correspond to integers, and
this is how we understand the expression
 \[
   \m(\ket{\psi} \mid N),
 \]
which is therefore nonzero only for elementary states $\ket{\psi}$.
(This is not our definition of quantum
universal probability or complexity, only a tool from classical complexity
theory helpful in its discussion.)

The quantum analog of a probability distribution is a density matrix,
a self-adjoint positive semidefinite operator with trace 1.
Just as with universal probability, let us allow operators with trace less
than 1, and call them \df{semi-density matrices}.

We call a sequence $A_{N}$ of operators, where $A_{N}$ is defined over
$\cH_{N}$, 
\df{lower semicomputable} if there is a double sequence of elementary
operators $A_{Nk}$ with the property that for each $N$, the sequence
$A_{Nk}$ is increasing and converges to $A_{N}$.

 \begin{lemma}\
 \begin{enumerate}
  \item
A computable sequence of operators is also lower semicomputable.
  \item
If $A_{N}$ is nonnegative then the elements of the sequence $A_{Nk}$ 
can be chosen nonnegative.
 \end{enumerate}
  \end{lemma}
 \begin{proof}
Both these statements are proved via standard approximations.
 \end{proof}

From now on, we suppress the index $N$ whenever it is not necessary to
point out its presence for clarity.

 \begin{theorem}\label{t.univ-dens}
There is a lower semicomputable semi-density matrix  $\bmu$ 
dominating all other such matrices in the sense that
for every other such matrix $\sg$ there is a constant $c_{\sg} > 0$
with $c_{\sg} \sg \le \bmu$.  
We have $\bmu \eqm \bmu'$ where
 \begin{equation}\label{e.mu-as-sum}
   \bmu' = \sum_{\ket{\psi}} \m(\ket{\psi}) \ket{\psi} \bra{\psi}.
 \end{equation}
Also
 \[
  \bmu \eqm \sum_{\nu} \m(\nu) \nu \eqm \sum_{P} \m(P) P/\dim P
 \]
where $\nu$ runs through all elementary
semi-density matrices and $P$ runs through all elementary projections.
 \end{theorem}
 \begin{proof}
The proof of the existence of $\bmu$
is completely analogous to the proof of Proposition~\ref{p.univ}.

To prove $\bmu \eqm \bmu'$, note first that the form of its definition
guarantees that $\bmu'$ is a lower semicomputable semi-density, and
therefore $\bmu' \lem \bmu$.
It remains to prove $\bmu \lem \bmu'$.
Since $\bmu$ is lower semicomputable, there is a nondecreasing sequence
$\bmu_{k}$ of elementary semi-density matrices
such that $\bmu = \lim_{k} \bmu_{k}$, with $\bmu_{0} = 0$.
For $k \ge 1$, let $\dg_{k} = \mu_{k} - \mu_{k-1}$.
Each of the nonnegative self-adjoint operators $\dg_{k}$ can be
represented as a sum
 \[
   \dg_{k} = \sum_{i=1}^{n}p_{ki}\ket{\phi_{ki}}\bra{\phi_{ki}}.
 \]
Thus, $\bmu = \sum_{ki} p_{ki} \ket{\phi_{ki}}\bra{\phi_{ki}}$, with a
computable sequence $p_{ki} \ge 0$, where $\sum_{k,i} p_{ki} < 1$.
The vectors $\ket{\phi_{ki}}$ and the values $p_{nk}$ can be chosen
elementary.
Noting $p_{ki} \lem \m(k,i) \lem \m(\ket{\phi_{ki}})$ finishes the proof.

The statement of
sum representations using projections and elementary density matrices is
weaker than the statement about $\bmu'$.
 \end{proof}

We will call $\bmu$ the \df{quantum universal (semi-) density matrix}.
Thus, the quantum universal probability of a quantum state $\ket{\psi}$ is given by
 \[
   \bra{\psi} \bmu \ket{\psi}.
 \]
A representation analogous to~\eqref{e.random-program}
holds also for the quantum universal probability
$\bmu$.
It is not necessary to introduce a quantum Turing machine in place
of a classical Turing machine, since instead of outputting an elementary
quantum state $\ket{\psi}$, 
we can just output the probabilities themselves, leaving the preparation of
the state itself to whatever device we want, which might as well be a
quantum Turing machine.
The output of $U(Z)$ classically is
a probability distribution over the set of
strings: string $x$ comes out with probability $\m(x)$.
When the outputs are quantum states $\ket{\phi}$ with probability
$\m(\ket{\phi})$, then
the relevant output is not the distribution
$\ket{\phi} \mapsto \m(\ket{\phi})$: 
by far not all this information is available.
The actual physical output is just the density matrix $\bmu'$ as given in
~\eqref{e.mu-as-sum}.
Thus, we take the projection associated with each possible output $\ket{\phi}$,
multiply it with its probability and add up all these terms.
Indeed, assume that $A$ is any self-adjoint operator expressing some
property.
The expected value of $A$ over $U(Z)$ is given by $\tr A \bmu'$.
In particular, suppose that for some quantum state $\ket{\psi}$ we measure
whether $U(Z)=\ket{\psi}$.
The measurement will give a ``yes'' answer with probability 
 \begin{align*}
  \sum_{\ket{\phi}} \m(\ket{\phi})|\braket{\phi}{\psi}|^{2}
    &= \sum_{\ket{\phi}} \m(\ket{\phi}) 
     \bra{\psi} (\ket{\phi} \bra{\phi}) \ket{\psi}
\\  &= \bra{\psi} \bmu' \ket{\psi} = \tr \ket{\psi} \bra{\psi} \bmu'.
 \end{align*}

These analogies suggest to us to define complexity also as a self-adjoint
operator: 
 \begin{equation}\label{e.complexity-operator}
   \bkg = -\log \bmu.
 \end{equation}

 \begin{proposition}
The operator function $A \mapsto \log A$ is monotonic.
 \end{proposition}
For a proof, see~\cite{Bhatia96}.
This implies the upper semicomputability of $(-\log \bmu)$.
For some readers 
to appreciate that the proposition is nontrivial, we mention that
for example $A \mapsto e^{A}$ is not monotonic (see the same
references).
We will also use the following theorem, which could be called the 
``quantum Jensen inequality'':

 \begin{proposition}
If $f(x)$ is a convex function in an interval $[a,b]$ containing the
eigenvalues of operator $A$ then for all $\ket{\psi}$ we have
 \begin{equation}~\label{e.quantum-Jensen-1}
  f(\bra{\psi} A \ket{\psi}) \le \bra{\psi} f(A) \ket{\psi}.
 \end{equation}
 \end{proposition}
 \begin{proof}
Easy, see~\cite{Wehrl78}.
 \end{proof}

This implies:

 \begin{lemma}
Let $f$ be  a function concave in the interval $[a, b]$,
and $\ket{\psi}$ a vector.
Then the function $A \mapsto \bra{\psi} f(A) \ket{\psi}$
is concave for self-adjoint operators $A$ 
whose spectrum is contained in $[a, b]$.
 \end{lemma}

We have now two alternative definitions for quantum complexity of a
pure state,
depending on the order of taking the logarithm and taking the expectation: 
 \begin{align}\label{e.Kpq}
   \Hl(\ket{\psi}) &= -\log \bra{\psi} \bmu \ket{\psi},
\\ \Hu(\ket{\psi}) &= -\bra{\psi} (\log \bmu) \ket{\psi} 
                    = \bra{\psi} \bkg \ket{\psi}.
 \end{align}
An inequality in one direction can be established between them easily:

 \begin{theorem}\label{t.log-prob-smaller}
 \[
  \Hl(\ket{\psi}) \le  \Hu(\ket{\psi}).
 \]
 \end{theorem}
 \begin{proof}
Use~\eqref{e.quantum-Jensen-1}.
 \end{proof}

The difference between the two quantities can be very large, as shown by
the following example.

 \begin{example}
Let $\ket{1},\dots,\ket{N}$ be the eigenvectors
of $\bmu$ ordered by decreasing eigenvalues $p_{i}$.
Then $p_{1}\eqm 1$ and $p_{N} \eqm N^{-1}$.
For vector $\ket{\psi} = 2^{-1/2}(\ket{1} + \ket{N})$ we have
 \begin{align*}
  \Hl(\ket{\psi}) &= 
  -\log \bra{\psi} \bmu \ket{\psi} = -\log (p_{1}/2 + p_{N}/2) \eqa 0,
\\  \Hu(\ket{\psi}) &= 
  \bra{\psi} \bkg \ket{\psi} = (-\log p_{1} -\log p_{N})/2 \eqa (\log N)/2.
 \end{align*}
 \end{example}

Which one of the two definitions is more appropriate?
We prefer $\Hu$ since we like the idea of a complexity operator;
however, in the present paper, we try to study both.

The complexity $\Kq$ introduced in~\cite{VitanyiThreeAppr00} can be
viewed as the formula resulting from $\Hl(\ket{\psi})$ when 
the sum in~\eqref{e.mu-as-sum} is replaced with supremum.
In classical algorithmic information theory, the result does not
change by more than a multiplicative constant after replacement,
but Theorem~\ref{t.Kq-lb} shows that it does in the quantum case.

 \begin{remark}
It seems natural to generalize $\Hu(\ket{\psi})$ and $\Hl(\ket{\psi})$ to density
matrices $\rho$ by
 \[
   \Hu(\rho) = \tr \bkg \rho,\quad \Hl(\rho) = -\log \tr \bmu \rho,
 \]
but we do not explore this path in the present paper, and are not even sure
that this is the right generalization.
 \end{remark}

\section{Properties of algorithmic entropy}\label{s.complexity-properties}

\subsection{Relation to classical description complexity}

It was one of the major attractions of the original Kolmogorov complexity
that it could be defined without reference to probability and then it could
be used to characterize randomness.
Unfortunately, we do not have any characterization, even to good
approximation, of $\Hu(\ket{\psi})$ or $\Hl(\ket{\psi})$
in terms avoiding probability.
As a generalization of classical complexity, it has the
properties of classical complexity in the original domain, just as
$\Kq$ and qubit complexity.

 \begin{theorem}
Let $\ket{1}, \ket{2}, \dotsc$ 
be a computable orthogonal sequence of states.  
Then for $H=\Hu$ or $\Hl$, we have
 \begin{equation}\label{e.rel-to-K}
  H(\ket{i}) \eqa K(i),
 \end{equation}
 where the constant in $\eqa$ depends on the definition of the sequence.
 \end{theorem}

 \begin{proof}
The function $f(i) = \bra{i} \bmu \ket{i}$ is lower
semicomputable with $\sum_{i} f(i) \le 1$, hence it is dominated by
$\m(i)$.
This shows $K(i) \lea \Hl(\ket{i})$.

On the other hand, the semi-density matrix
$\rho = \sum_{i} \m(i) \ket{i} \bra{i}$ is lower semicomputable, so 
$\rho \lem \bmu$, $-\log \rho \gea \bkg$, hence
 \[
 K(i) = \bra{i}(-\log \rho) \ket{i}
      \gea \bra{i} \bkg \ket{i} = \Hu(\ket{i}).
 \]
 \end{proof}

\subsection{Upper and lower bounds in terms of small simple subspaces}

The simple upper bound follows immediately from the domination property
of universal probability.

 \begin{theorem}
Assume that $\ket{\psi} \in \cH_{N}$.  Then
 \[
   \bkg \lea (\log N) \bid.
 \]
In particular, if $\ket{\psi} \in \cQ_{n}$ then $\Hu(\ket{\psi}) \lea n$.
 \end{theorem}
 \begin{proof}
Let $\rho = N^{-1}\bid$, then $\rho \lem \bmu$, hence
$\bkg \lea (\log N) \bid$.
 \end{proof}

 \begin{remark}
$N$ is an implicit parameter here, so it is more correct to write
$\bkg(\cdot \mid N) \lea (\log N) \bid$.
We do not have any general definition of quantum conditional complexity
(just as no generally accepted notion of quantum conditional entropy is
known), but conditioning on a classical parameter is not problematic.
 \end{remark}

There is a more general theorem for classical complexity.
For a finite set $A$ let $K(A)$ be the length of the shortest program
needed to enumerate the elements of $A$.
Then for all $x\in A$ we have
 \[
   K(x) \lea K(A) + \log \# A + 2\log \# A.
 \]
What may correspond to a simple finite set $A$ is a
projector $P$ that is lower semicomputable as a nonnegative operator.
What corresponds to $\# A$ is the dimension $\tr P$ of the subspace to
which $P$ projects.
What corresponds to $x \in A$ is measuring the angle between 
$\ket{\psi}$ and the space to which $P$ projects.

 \begin{theorem}\label{t.in-small-set}
Let $P$ be a lower semicomputable projection with $d = \tr P$.
We have
 \begin{align}\label{e.Hl-in-small-set}
  \Hl(\ket{\psi}) &\lea K(P) + \log d - \log \bra{\psi} P \ket{\psi},
\\\label{e.Hu-in-small-set}
  \Hu(\ket{\psi}) &\lea K(P) + \log d + (1 - \bra{\psi}P\ket{\psi}) \log N. 
 \end{align}
 \end{theorem}
 \begin{proof}
Let $\rho$ be the semi-density matrix 
 \[
   \half(\frac{P}{d} + \frac{\bid - P}{N}) = \half(\bid/N + P(1/d - 1/N))
 \]
From the first form, it can be seen that it is semi-density, from the
second form, it can be seen that it is lower semicomputable.
By Theorem~\ref{t.univ-dens}, we have $2^{K(\rho)}\rho \lem \bmu$.
Since $K(\rho) \eqa K(P)$, we have
 \[
  \Hl(\ket{\psi}) = -\log \bra{\psi} \bmu \ket{\psi}
              \lea K(P) + \log \bra{\psi} (P/d) \ket{\psi}
  \eqa K(P) + \log d - \log \bra{\psi} P \ket{\psi}.
 \]
On the other hand,
 \begin{align*}
  \Hu(\ket{\psi}) 
    &= \bra{\psi} (-\log \bmu) \ket{\psi}
\\  &\lea K(P) + \bra{\psi} P \ket{\psi} \log d
     + (1 - \bra{\psi}P\ket{\psi})\log N.
 \end{align*}
 \end{proof}

This theorem points out again the difference between $\Hl$ and
$\Hu$.
If $\ket{\psi}$ has a small angle with a small-dimensional subspace this
makes $\Hl(\ket{\psi})$ small.
For $\Hu(\ket{\psi})$, the size of the angle gets
multiplied by $\log N$, so if nothing more is known about $\ket{\psi}$ then
not only the dimension of $P$ counts
but also the dimension of the whole space we are in.

Above, we defined what it means for a program to recursively ``enumerate a
subspace'' by saying that it approximates the projector from below
as a nonnegative operator: call this ``weak enumeration''.
There is a simpler possible definition: let the program just list a
sequence of orthogonal vectors that generate the subspace: call this
``strong enumeration''.

\begin{remarks}\
 \begin{enumerate}
  \item The rest of the paper makes no use of the discussion of strong
and week enumeration, so this part can be skipped.
  \item What is important is not only
that the sequence of vectors in question can be enumerated, since this is
in some sense trivially true for any finite sequence of elementary vectors.
A recursively enumerable finite-dimensional subspace is
always elementary.
What matters is that the enumeration is done with a short program (which
can use the dimension $N$ as input).
Without this remark, there is clearly no difference between an elementary
subspace and a strongly enumerable one.
 \end{enumerate}
\end{remarks}

 \begin{proposition}
The strong and weak kinds of enumeration of a subspace are equivalent. 
In other words, there is a program of length $k$ enumerating a subspace in
the weak sense if and only if there is a program of length $\eqa k$
enumerating it in the strong sense.
 \end{proposition}

 \begin{proof}
Given a strong enumeration $\ket{\phi_{1}}, \ket{\phi_{2}}, \dotsc$, the
sum $\sum_{i} \ket{\phi_{i}} \bra{\phi_{i}}$ clearly defines the projector
in a form 
from which the possiblity of approximating it from below is seen.

Assume now that $P$ is a projector and $\rho_{1} \le \rho_{2} \le \dotsb$
is a sequence of elementary nonnegative operators approximating it.

Note that for a nonnegative operator $A$, we have 
$\bra{\psi}A\ket{\psi} = 0$ iff $A\ket{\psi} = 0$.
Now for any of the $\rho_{i}$, and any vector $\ket{\psi}$, if $P\ket{\psi}=0$
then $\bra{\psi}P\ket{\psi} = 0$, which implies 
$\bra{\psi}\rho_{i}\ket{\psi} = 0$ and thus $\rho_{i} \ket{\psi} = 0$.
Hence the kernel of $\rho_{i}$ contains the kernel of $P$ and hence the
space of eigenvectors of $\rho_{i}$ with nonnegative eigenvalues is
contained in $P\cH$.
This shows that from $\rho_{i}$, $i=1,2,\dotsc$ we will be able to build up
a sequence $\ket{\phi_{1}}, \ket{\phi_{2}}, \dotsc$
of orthogonal vectors spanning $P\cH$.
 \end{proof}

Theorem~\ref{t.H-lb} below is
analogous to the simple lower bound on classical description complexity.
That lower bound
says that the number of objects $x$ with $K(x) < k$ is at most $2^{k}$.
What corresponds here to ``number of objects'' is dimension, and the
statement is approximate: 
if $\ket{\psi}$ has complexity $< k$ then it is within a small angle
from a certain fixed $2^{k+1}$-dimensional space.
The angle is really small for $\Hu$; it is not so small for $\Hl$
but it is still small enough that the whole
domain within that angle makes up only a small portion of the Hilbert
space.

Let $\ket{u_{1}}, \ket{u_{2}}, \dotsc$ be the sequence of eigenvectors of
$\bmu$  with eigenvalues $\mu_{1} \ge \mu_{2} \ge \dotsb$.
(Since our space is finite-dimensional, the sequence exists.)
Let $\kg_{i} = -\log \mu_{i}$.
Let $E_{k}$ be the projector to
the subspace generated by $\ket{u_{1}},\dots,\ket{u_{k}}$.

 \begin{remark}
The universal density matrix $\bmu$
is an object with an impressive invariance
property: for any other universal density matrix $\nu$ we have
$\nu \eqm \bmu$.
On the other hand, the
individual eigenvectors $\ket{u_{i}}$ probably do not have any
invariant significance.
It is currently not clear whether even the projectors $E_{k}$ enjoy
any approximate invariance property.
 \end{remark}

 \begin{theorem}[Lower bounds]\label{t.H-lb}\
Let $\ket{\psi}$ be any vector and let $\lg > 1$.
If $\Hu(\ket{\psi}) < k$ then we have
 \begin{equation}\label{e.Hu-lb}
  \bra{\psi} E_{2^{\lg k}} \ket{\psi} > 1 - 1/\lg.
 \end{equation}
If $\Hl(\ket{\psi}) < k$ then we have
 \begin{equation}\label{e.Hl-lb}
  \bra{\psi} E_{\lg 2^{k}} \ket{\psi} > 2^{-k}(1 - 1/\lg).
 \end{equation}
 \end{theorem}
 \begin{proof}
Assume $\Hu(\ket{\psi}) < k$ and
expand $\ket{\psi}$ in the basis $\{\ket{u_{i}}\}$ 
as $\ket{\psi} = \sum_{i} c_{i} \ket{u_{i}}$.
By the assumption, we have $\sum_{i} \kg_{i} |c_{i}|^{2} < k$.
Let $m$ be the first $i$ with $\kg_{i} > \lg k$.
Since $\sum_{i} 2^{-\kg_{i}} < 1$  we have $m \le 2^{\lg k}$.
Also,
 \[
  \lg k  \sum_{i \ge m} |c_{i}|^{2} <
  \sum_{i \ge m} \kg_{i} |c_{i}|^{2} < k,
 \]
hence $\sum_{i \ge m} |c_{i}|^{2} < 1/\lg$, which proves~\eqref{e.Hu-lb}.

Now assume $\Hl(\ket{\psi}) < k$, then 
we have $\sum_{i} \mu_{i} |c_{i}|^{2} \ge 2^{-k}$.
Let $m$ be the first $i$ with $\mu_{i} < 2^{-k} / \lg$.
Since $\sum_{i} \mu_{i} < 1$  we have $m \le 2^{k} \lg$.
Also,
 \[
  \sum_{i \ge m} \mu_{i} |c_{i}|^{2} 
     < 2^{-k}/\lg \sum_{i} |c_{i}|^{2} = 2^{-k} / \lg, 
 \]
hence 
 \begin{equation}\label{e.Hl-lb-meat}
 \begin{split}
  \bra{\psi} E_{m} \ket{\psi} &= \sum_{i < m} |c_{i}|^{2}
                          > \sum_{i < m} \mu_{i} |c_{i}|^{2}
                          \ge 2^{-k} - \sum_{i \ge m} \mu_{i} |c_{i}|^{2} 
\\                       &> 2^{-k}(1 - 1/\lg).
 \end{split}
 \end{equation}
\end{proof}

The defect of this theorem is that the operators $E_{k}$ are uncomputable. 
I do not know whether the above properties can be claimed for some
lower semicomputable operators $F_{k}$.

\subsection{Quantum description complexities}

\subsubsection{Vit\'anyi's complexity}

Theorem~\ref{t.Kq-Hl}
says that the complexity $\Kq$ from ~\cite{VitanyiThreeAppr00},
(defined in Section~\ref{s.intr}) is not too much larger than $\Hl$, so
we do not lose too much in replacing the sum~\eqref{e.mu-as-sum} with a
supremum: if the sum is $> 2^{-k}$ 
then the supremum is $> 2^{-4k}/k^{2}$.

 \begin{theorem}[Relation to $\Kq$]\label{t.Kq-Hl}
 \begin{equation}\label{e.Kq-Hl}
 \Hl \lea \Kq \lea 4 \Hl + 2\log \Hl.
 \end{equation}
 \end{theorem}

 \begin{proof}
We start from the end of the proof of Theorem~\ref{t.H-lb}.
We use~\eqref{e.Hl-lb-meat} with
$\lg = 2$, and note that one term, say, $|c_{r}|^{2}$ of the sum
$\sum_{i \le m} |c_{i}|^{2}$ must be at least $2^{-2k-2}$.
We would be done if we could upperbound $K(\ket{u_{r}})$ appropriately.
It would seem that $K(\ket{u_{r}})$ can be bounded approximately by $k$ since
$m \le 2^{k+1}$.
But unfortunately, neither the vectors $\ket{u_{i}}$ nor their sequence are
computable; so, an approximation is needed.
Let $r$ be the largest binary number of length
$\le k$ smaller than $\tr \bmu$.
Then there is a program $p$ of length $\le k + 2\log k$
computing a lower approximation $\hat{\bmu}$ of $\bmu$ such that
$\tr \bmu - \tr \hat{\bmu} \le 2^{-k}$.
Indeed, let $p$ specify the binary digits of $r$ and then compute an
approximation of $\tr \bmu$ that exceeds $r$.

The condition $\bra{\psi}\bmu\ket{\psi} \ge 2^{-k}$ implies
$\bra{\psi}\hat{\bmu}\ket{\psi} \ge 2^{-k+1}$.
We can now proceed with $\hat{\bmu}$ as with $\bmu$.
We compute eigenvectors $\ket{\hat u_{i}}$ for $\hat{\bmu}$, and 
find an elementary vector $\ket{\hat u_{r}}$ with
 \[
                   K(\ket{\hat u_{r}}) \lea 2 k + 2 \log k,\quad
  |\braket{\psi}{\hat u_{r}}|^{2} \gem 2^{-2k}.
 \]
The extra $k + 2 \log k$ in $K(\ket{\hat u_{r}})$ is coming from the
program $p$ above.
 \end{proof}

\subsubsection{Qubit complexity}

Let us define the qubit complexity introduced
in~\cite{BerthDamLaplante00}.
We refer to that paper for further references 
on quantum Turing machines and detailed specifications
of the quantum Turing machine used.
Our machine starts from an input (on the input tape) consisting
of a qubit program and a rational number $\eps > 0$.
On the output tape, an output appears, preceded by a 0/1 symbol telling whether
the machine is considered halted.
The halting symbol as well as the content of the output tape does not
change after the halting symbol turns 1.
(The input tape, which is also the work tape, keeps changing.)
We can assume that input and output strings of different lengths can always
be padded to the same length at the end by 0's, or if this is inconvenient,
by some special ``blank'', or ``vacuum'' symbol.
The input of the machine is a density matrix $\rho$.
For any segment of some length $n$ of the output, and any given time $t$
there is a completely positive operator $\Phi_{k,t}$ such that 
the $n$ symbols of the output at time $t$ are described by a density matrix
$\sg = \Phi_{k,t} \rho$.
We only want to consider the output state when the machine halted.
If $H$ is a projection to the set of those states then the semi-density
matrix $H\sg H$ is the output we are interested in.
The operation $\Psi_{n,t} : \rho \mapsto H\sg H$ is a completely positive
operator but it is not trace-preserving, it may decrease the trace.
It is also monotonically increasing in $t$.

For a state $\ket{\psi}$, let 
$\QC^{\eps}(\ket{\psi})$ be the length $k$ of the smallest qubit program (an
arbitrary state in $\cQ_{k}$, or more precisely
the density matrix corresponding to this
pure state) which, when given as input along with $\eps$, results in an
output density matrix $\sg$ with $\bra{\psi}\sg\ket{\psi} \ge 1 - \eps$.
The paper~\cite{BerthDamLaplante00} shows that this quantity has the
same machine-independence properties as Kolmogorov complexity, so we also
assume that a suitable universal quantum Turing machine has been fixed.
For the following theorem, we will compute complexities of strings
in $\cH_{N}=\cQ_{n}$, so~$N = 2^{n}$.

 \begin{lemma}\label{l.approx}
If for a semi-density matrix $\rho$ and a state $\ket{\psi}$ we have
$\bra{\psi} \rho \ket{\psi} \ge 1 - \eps$ and $\rho$ has the eigenvalue
decomposition $\sum_{i} p_{i} \ket{i}\bra{i}$ where 
$p_{1} \ge p_{2} \ge \dotsb$, then
 \[
   p_{1} \ge 1 - \eps,\quad |\braket{1}{\psi}|^{2} \ge 1 - 2\eps.
 \]
 \end{lemma}
 \begin{proof}
Let $c_{i} = \braket{i}{\psi}$, then
$\bra{\psi}\rho\ket{\psi} = \sum_{i} p_{i} |c_{i}^{2}| \ge 1 -\eps$.
Hence $p_{1} \ge 1 - \eps$, therefore
 \[
  |c_{1}|^{2} + \eps \ge \sum_{i} p_{i} |c_{i}^{2}| \ge 1 - \eps,
 \]
giving~$|c_{1}^{2}| \ge 1 - 2\eps$.
 \end{proof}

 \begin{theorem}\label{t.H-less-QC}
For $\eps < 0.5$, if $\QC^{\eps}(\ket{\psi}) \le k$ then
 \[
   \Hu(\ket{\psi}) \lea k + K(k) + 2\eps n.
 \]
 \end{theorem}

 \begin{proof}
For each $k$, let $I_{k}$ be the projection to the space $\cQ_{k}$ of
$k$-length inputs.
The operator
 \[
   \lg = \sum_{k} \m(k) 2^{-k} I_{k}
 \]
is a semicomputable semi-density matrix on the set of all inputs.
For each time $t$, the semi-density matrix $\Psi_{n,t} \lg$ 
is semicomputable.
As it is increasing in $t$, the limit $\nu = \lim_{t} \Psi_{n,t} \lg$ is a
semicomputable semi-density matrix, and therefore $\nu \lem \bmu$.
Let $\ket{\phi} \in \cQ_{k}$, then $\ket{\phi}\bra{\phi} \le I_{k}$, hence
$\m(k) 2^{-k} \ket{\phi}\bra{\phi} \le \lg$, hence for each $t$ we have
 \[
  \m(k) 2^{-k} \Psi_{t,k} \ket{\psi}\bra{\psi} \le \nu \lem \bmu.
 \]
Since also $2^{-n}I_{n} \lem \bmu$, we can assert, with 
$\rho_{t,k} = \Psi_{t,k} \ket{\phi}\bra{\phi}$, that 
 \[
   \sg = \m(k) 2^{-k} \rho_{t,k} + 2^{-n} I_{n} \lem \bmu.
 \]
Assume that $\bra{\psi}\rho_{t,k}\ket{\psi} \ge 1 - \eps$.
Then by Lemma~\ref{l.approx}, if $\rho_{t,k}$ has the eigenvalue
decomposition $\sum_{i} p_{i} \ket{i}\bra{i}$ then
$p_{1} \ge 1 - \eps$ and $|\braket{1}{\psi}|^{2} \ge 1 - 2\eps$.
The matrix $(-\log \sg)$ can be written as
 \[
 - \sum_{i} \log (\m(k) 2^{-k} p_{i} + 2^{-n}) \ket{i}\bra{i}.
 \]
  Hence, with $c_{i} = \braket{i}{\psi}$, and using
Lemma~\ref{l.approx} and $\eps < 0.5$
 \begin{align*}
   -\bra{\psi}\log \bmu \ket{\psi} &\lea -\bra{\psi}\log \sg \ket{\psi} 
\\  &= \sum_{i} \log (\m(k) 2^{-k} p_{i} + 2^{-n})|c_{i}|^{2}
\\  &\le k + K(k) + \log(1-\eps) + 2 \eps n.
 \end{align*}
In the last inequality, the first two terms come from the first term of the
previous sum, while $2\eps n$ comes from the rest of the terms.
 \end{proof}

Using the definitions of~\cite{BerthDamLaplante00}, we write
$\QC(\ket{\psi}) \le k$ 
if there is a $\ket{\phi}$ such that for all $\eps$ of the form $1/m$, when
$\ket{\phi}$ is given as input along with $\eps$, we get an output density
matrix $\sg$ with $\bra{\psi}\sg\ket{\psi} \ge 1 - \eps$.
The above theorem implies that in this case,
 \begin{equation}\label{e.H-less-QC}
   \Hu(\ket{\psi}) \lea k + K(k).
 \end{equation}
Let $x$ be a bit string, then we know from~\eqref{e.rel-to-K} that 
 \begin{equation}\label{e.rel-to-K-x}
  \Hu(\ket{x}) \eqa K(x).
 \end{equation}
It has been shown in~\cite{BerthDamLaplante00} that 
$\QC(\ket{x}) \lea C(x)$ 
where $C(x)$ is the (not prefix-free) Kolmogorov complexity.
We can show directly that also $C(x) \lea \QC(\ket{x})$, but we will not do
it in this paper.
It follows from~\eqref{e.H-less-QC} and~\eqref{e.rel-to-K-x} that
$K(x) \eqa \Hu(\ket{x}) \lea \QC(\ket{x}) + K(\QC(\ket{x}))$.
This is in some way stronger, since another interesting quantity,
$\Hu(\ket{x})$ is interpolated, and in another way it seems slightly 
weaker.
But only very slightly, since one can bound $K(x)$ by $C(x)$
in general only via $K(x) \lea C(x) + K(C(x))$.

Just as we obtained an upper bound on $\Kq$ using~\eqref{e.Hl-lb} combined
with an approximation of the uncomputable $\bmu$, we may hope to obtain an
upper bound on $\QC$ using~\eqref{e.Hu-lb} combined with a suitable
approximation of the uncomputable $\bmu$ or $(-\log\bmu)$.
But we did not find an approximation in this case 
for a reasonable price in complexity: the best we can say 
replaces $\Hu(\ket{\psi})$ with $\bra{\psi}(-\log \mu)\ket{\psi}$ for any
computable density matrix $\mu$.
Or, we can upperbound not $\QC(\ket{\psi})$ but $\QC(\ket{\psi} \mid \chi)$
where $\chi$ is an 
encoding of the halting problem into a suitable infinite binary string.
The concept of an oracle quantum computation with a read-only classical
oracle tape presents no difficulties.

 \begin{theorem}\label{t.QClessH}
For each rational $\eps$ and 
any computable density matrix $\mu$ we have 
 \[
   \QC^{\eps}(\ket{\psi}) \lea \bra{\psi}(-\log \mu)\ket{\psi}/\eps + K(\mu).
 \]
Similarly, 
 \[
   \QC^{\eps}(\ket{\psi} \mid \chi) \lea \Hu(\ket{\psi})/\eps.
 \]
 \end{theorem}
 \begin{proof}
For the second inequality, we
can use~\eqref{e.Hu-lb} with $k = \Hu(\ket{\psi})$ and $\lg = 1/\eps$.
The oracle $\chi$ allows us to compute the space $E_{2^{\lg k}}$ with
arbitrary precision.
Then our quantum Turning machine can simply map the space of
$\lg k$-length qubit strings into the (approximate) $E_{2^{\lg k}}$.

Similarly, for the first inequality,
if $\mu$ is computable then we can compute the subspaces
corresponding to $E_{2^{\lg k}}$ with arbitrary precision.
 \end{proof}

\subsection{Invariance under computable transformations}

 \begin{theorem}
Let $U$ be any computable unitary transformation.
Then we have
 \[
   \Hu(U \ket{\psi}) \eqa \Hu(\ket{\psi}),\quad 
   \Hl(U \ket{\psi}) \eqa \Hl(\ket{\psi}).
 \]
 \end{theorem}
 \begin{proof}
Straightforward.
 \end{proof}

This theorem needs to be generalized: it should be understood how
complexity changes under a completely positive operator.

\section{Complexity and entropy}

In classical algorithmic information theory, if $\rho$ is a discrete
computable probability distribution then its entropy is equal, 
to a good approximation, to the average complexity.
In the quantum case, entropy is defined as
 \[
   S(\rho) = -\tr \rho \log \rho.
 \]
There is a quantity corresponding to the Kullback information distance, and 
called \df{relative entropy} in~\cite{Wehrl78}: it is defined as
 \[
   S(\rho \prl \sg) = \tr \rho(\log \rho - \log \sg),
 \]
where $\rho$ and $\sg$ are density matrices.

 \begin{proposition} 
 \begin{equation}\label{e.quant-Kullback-pos}
   S(\rho \prl \sg) \ge 0.
 \end{equation}
 \end{proposition}
 \begin{proof}
See~\cite{Wehrl78}.
 \end{proof}

The following theorem can be interpreted as saying that entropy is equal to
average complexity:
 \begin{theorem}
For any lower semicomputable semi-density matrix $\rho$ we have
 \begin{equation}\label{e.quant-average-compl}
  S(\rho) \eqa \tr \rho \bkg
 \end{equation}
 \end{theorem}
 \begin{proof} 
Let $\Og = \tr \bmu$, then $\sg = \bmu/\Og$ is a density matrix, and
hence by~\eqref{e.quant-Kullback-pos}, $S(\rho \prl \sg) \ge 0$.
It follows that $S(\rho) \lea \tr \rho \bkg$.
 
On the other hand, since $\rho \lem \bmu$, the monotonicity of logarithm
gives $\bkg \lea -\log \rho$ which gives the other inequality.
 \end{proof}

For what follows the following property of logarithm is useful:

 \begin{lemma}
If $A$ and $B$ are nonnegative operators over $X$ and $Y$ respectively,
then 
 \begin{equation}\label{e.log-otimes}
  \log A \otimes B = (\log A) \otimes \bid_{Y} + \bid_{X} \otimes (\log B).
 \end{equation}
 \end{lemma}
 \begin{proof}
Direct computation.
 \end{proof}

Some properties of complexity that can be deduced from its universal
probability formulation will carry over to the quantum form.
As an example, take subadditivity:
 \[
   K(x, y) \lea K(x) + K(y).
 \]
What corresponds to this in the quantum formulation is the following:

 \begin{theorem}[Subadditivity]
We have
 \begin{equation}\label{e.product-subadd}
  \bmu_{X} \otimes \bmu_{Y} \lem \bmu_{XY}.
 \end{equation}
For $\ket{\phi}, \ket{\psi} \in \cH_{N}$ and $H=\Hu$ or $\Hl$ we have
 \begin{equation}\label{e.quant-compl-subadd}
   H(\ket{\phi}  \ket{\psi}) \lea H(\ket{\phi}) + H(\ket{\psi}).
 \end{equation}
 \end{theorem}
 \begin{proof}
The density matrix $\bmu_{X} \otimes \bmu_{Y}$ 
over the space $\cH_{XY} = \cH_{X} \otimes \cH_{Y}$ is
lower semicomputable, therefore~\eqref{e.product-subadd} follows.
Hence 
 \begin{align*}
 (\bra{\phi}\bmu_{X} \ket{\phi}) (\bra{\psi}\bmu_{Y} \ket{\psi})
   &= \bra{\phi}  \bra{\psi}(\bmu_{X} \otimes \bmu_{Y})
      \ket{\phi}  \ket{\psi} 
\\ &\lem
 \bra{\phi}  \bra{\psi} \bmu_{XY} \ket{\phi}  \ket{\psi}.
 \end{align*}
which gives~\eqref{e.quant-compl-subadd} for $H = \Hl$.
For $H = \Hu$ note that by the monotonicity of logarithm, identity
~\eqref{e.log-otimes} and~\eqref{e.product-subadd} implies
 \begin{align*}
   (\log \bmu_{X}) \otimes \bid_{Y} + 
   \bid_{X} \otimes (\log \bmu_{Y}) 
  &= \log \bmu_{X} \otimes \bmu_{Y} \lea \log \bmu_{XY}.
 \end{align*}
Taking the expectation (multiplying by $\bra{\psi}$ on left and
$\ket{\psi}$ on right) gives the desired result.
 \end{proof}

The analogous
subadditivity property also holds for the quantum entropy $S(\rho)$.

For classical complexity we have $K(x) \lea K(x,y)$,
and the corresponding property also holds for classical entropy.
This monotonicity property can also be proved for quantum complexity.

 \begin{theorem}[Monotonicity]\label{t.quant-compl-mon}
We have
 \begin{align}\label{e.aprob-mon}
   \tr_{Y} \bmu_{XY} &\eqm \bmu_{X},
\\\label{e.bkg-mult-bid}
           \bkg_{XY} &\gea \bkg_{X} \otimes \bid_{Y}.
 \end{align}
For $\ket{\phi}, \ket{\psi} \in \cH_{N}$, and $H = \Hu$ or $\Hl$ we have
 \begin{equation}\label{e.quant-compl-mon}
   H(\ket{\phi}) \lea H(\ket{\phi}  \ket{\psi}).
 \end{equation}
 \end{theorem}
 \begin{proof}
Let $\rho_{X} = \tr_{Y} \bmu_{XY}$.
Then $\rho_{X}$ is a semicomputable semi-density matrix over $\cH_{X}$ and
thus $\rho_{X} \lem \bmu_{X}$.
At the same time, for any fixed vector $\ket{\psi}$, the matrix 
$\sg_{XY} = \bmu_{X} \otimes \ket{\psi}\bra{\psi}$ is a lower
semicomputable semi-density matrix, hence $\bmu_{XY} \gem \sg_{XY}$.
Taking the partial trace gives 
 \[
   \bmu_{X} = \tr_{Y} \sg_{XY} \lem \tr_{Y} \bmu_{XY} = \rho_{X}.
 \]
This proves~\eqref{e.aprob-mon}, which implies the inequality for $\Hl$.

Let $\{\ket{\psi_{i}}\}$ be any orthogonal basis of $\cH_{Y}$ with
$\ket{\psi_{1}} = \ket{\psi}$.
Then we have
 \[
  \begin{split}
  \bra{\phi}\bra{\psi}\bmu_{X} \otimes \bid_{Y} \ket{\phi} \ket{\psi}
      &= \bra{\phi} \bmu_{X} \ket{\phi} 
\\ 
      &\eqm \bra{\phi} \tr_{Y}\bmu_{XY} \ket{\phi} 
  = \sum_{i} 
\bra{\phi}  \bra{\psi_{i}} \bmu_{XY} \ket{\phi}\ket{\psi_{i}}
  \ge \bra{\phi}\bra{\psi} \bmu_{XY} \ket{\phi}\ket{\psi},
  \end{split}
 \]
which proves $\bmu_{X} \otimes \bid_{Y} \gem \bmu_{XY}$.
Taking logarithms and noting that $\log \bid_{Y} = 0$, we 
get~\eqref{e.bkg-mult-bid} which proves the inequality for $\Hu$. 
 \end{proof}

The quantum entropy analog of this monotonicity fails in a spectacular way.
It is not true in general that $S(\rho_{X}) \le S(\rho_{XY})$.
Indeed, $\rho_{XY}$ could be the density matrix of a pure state, and then
$S(\rho_{XY}) = 0$.
At the same time, if this pure state is an entangled state, a state
that cannot be represented in the form of $\ket{\phi}  \ket{\psi}$, only as
the linear combination of such states, then $S(\rho_{X}) > 0$.
This paradox does not contradict to the possibility that entropy is
``average complexity''.
It just reminds us that Theorem~\ref{t.quant-compl-mon} says nothing about
entangled states.
An entangled state can be simple even if it is a big sum, but in this case
it will contain a lot of complex components.

\section{The cloning problem}

\subsection{Maximal complexity of cloned states}

For classical description complexity, the relation 
 \[
  K(x, x) \eqa K(x)
 \]
holds and is to be expected: once we have $x$ we can copy it and get
the pair $(x,x)$.
But there is a ``no cloning theorem''~\cite{PeresBook95}
in quantum mechanics saying that there is 
no physical way to get $\ket{\psi}  \ket{\psi}$ from $\ket{\psi}$.
It is interesting to see that a much stronger form of this theorem
also holds, saying that sometimes
$\Hu(\ket{\psi}  \ket{\psi})$ is much larger
than $\Hu(\ket{\psi})$ (of course, at most twice as large).
Moreover, we can determine the maximum complexity of states of the form
$\ket{\psi}^{\otimes k}$.
Our results in this are very similar in form to those of
~\cite{BerthDamLaplante00}, and the proof method is also similar.

For $\ket{\psi} \in \cH_{N}$, let $\ket{\psi}^{\otimes m}$ denote the $m$-fold
tensor product of $\ket{\psi}$ with itself, an element of $\cH^{\otimes m}$.

Let
 \[
   \cS_{N,m} = \cH^{\vee m} \sbs \cH_{N}^{\otimes m}
 \]
be the subspace of elements of $\cH_{N}^{\otimes m}$ invariant
under the orthogonal 
transformations arising from the permutations
 \[
  \ket{\phi_{1}}\dots\ket{\phi_{m}} \mapsto
  \ket{\phi_{\pi(1)}}\dots\ket{\phi_{\pi(m)}}.
 \]
 \begin{lemma}[see~\cite{WeylClassical46}]\label{l.symm-irred}\
 \begin{enumerate}
  \item $\dim \cS_{N,m} = \binom{m + N - 1}{m}$.
  \item  $\cS_{N,m}$ is invariant under unitary 
transformations of the form $U^{\otimes m}$.
  \item If a density matrix over $\cS_{N,m}$ commutes with all such
transformations then it is a multiple of unity.
 \end{enumerate}
 \end{lemma}

Let 
 \begin{equation}\label{e.CNm}
   \ol C_{N, m} = \max_{\ket{\psi} \in \cH_{N}} \Hu(\ket{\psi}^{\otimes m}),
 \end{equation}
and let $\ul C_{N,m}$ be defined the same way with $\Hl$ in place of
 $\Hu$.

 \begin{theorem}\label{t.cloning}
We have
 \begin{align*}
          \ol C_{N,m}  &\lea  K(m) + \log \binom{m + N - 1}{m},
\\        \ul C_{N,m}  &\ge  \log \binom{m + N - 1}{m}.
 \end{align*}
 \end{theorem}
 \begin{proof}
The upper bound follows from the fact that $\ket{\psi}\in \cS_{N,m}$ and from
~\eqref{e.Hu-in-small-set}.

For simplicity, let us write for the moment, $\ket{\psi}^{m} = \ket{\psi}^{\otimes m}$.
For the lower bound, let us first set $c = \ul C_{N,m}$.
We have
 \begin{equation}\label{e.cloning1}
 \tr \bmu \ket{\psi}^{m}\bra{\psi}^{m} = 
\bra{\psi}^{m} \bmu \ket{\psi}^{m} \ge 2^{-c}
 \end{equation}
for all states $\ket{\psi}\in \cH_{N}$.
Let $P_{S}$ be the projection to $\cS_{N,m}$.
Let $\Lambda$ be the uniform distribution on the unit sphere in $\cH_{N}$.
Then
 \[
   \rho = \int \ket{\psi}^{m}\bra{\psi}^{m}\,d\Lambda
 \]
is a density matrix over $\cS_{N,m}$.
It commutes with all unitary transformations of the form
$U^{\otimes m}$, and therefore according to Lemma~\ref{l.symm-irred}, 
 \[
   \rho = \binom{m + N - 1}{m}^{-1}P_{S}.
 \]
Integrating~\eqref{e.cloning1} by $d\Lambda$ we get
 \[
 2^{-c} \le \tr\bmu\rho 
        = \binom{m + N - 1}{m}^{-1} \tr \bmu P_{S}
       \le \binom{m + N - 1}{m}^{-1}.
 \]
Taking negative logarithm, we get the lower bound on $\ul C$.
 \end{proof}

\subsection{An algebraic consequence}

This subsection says nothing new about quantum complexities, it only
draws some technical inferences from the previous subsection.

The problem of estimating $\Hu(\ket{\psi}  \ket{\psi})$ can be reformulated
into an algebraic problem for which we are not aware of any previous
solution.
The results obtained above solve the problem: maybe such a solution will
also have some independent interest.
For any $N\times N$ matrix $A$, let
 \[
  u(A) = \frac{\nm{A^{\dgr}A}}{\tr A^{\dgr}A} =
 \max_{i} \frac{\ag_{i}}{\sum_{j} \ag_{j}}
 \]
where $\ag_{j}$ are the eigenvalues of $A^{\dgr}A$.
The function $u(A)$ measures the ``unevenness'' of 
the distribution of eigenvalues of $A^{\dgr}A$.
It can vary between $1/N$ for $A=\bid$ and 1 (when $A^{\dgr}A$ has rank
1).
For a subspace $F$ of the vector space of symmetric (not necessarily
self-adjoint!) matrices, let $u(F) = \max_{A \in F} u(A)$.
Let $N' = N(N+1)/2$.
For $0 < d < N'$, we are interested in the quantity
 \[
 u(d, N) = \min\setof{u(F) : \dim F \ge d}.
 \]

 \begin{theorem}\label{t.algeb}
  We have $u(d, N) \ge d / N'$.
 \end{theorem}
 \begin{remark} This theorem has been strengthened from its preprint
version.
 \end{remark}

Before the proof, we give some lemmas setting up the connection with
cloning.

 \begin{lemma}\label{l.uA}
Let $A$ be a symmetric $N\times N$ matrix $(a_{ij})$ and let
 \[
  \ag  = \sum_{ij} a^{*}_{ij}\, \ket{\bg_{i}}  \ket{\bg_{j}}.
 \]
Then
 \begin{equation}\label{e.uA}
  \sup_{\ket{\phi} \in \cH_{N}}
 |\bra{\ag}(\ket{\phi}\ket{\phi})|^{2} = u(A).
 \end{equation}
 \end{lemma}

 \begin{proof}
We can restrict ourselves to matrices $A$ 
with $\tr A^{\dgr} A = \braket{\ag}{\ag} = 1$.
Then with $\ket{\psi} = \ket{\phi}\ket{\phi}$,
$\ket{\phi} = \sum_{i} x_{i} \ket{\bg_{i}}$,
 \[
   |\braket{\ag}{\psi}|^{2} 
   = |\sum_{ij} a_{ij} x_{i} x_{j}|^{2} = |x^{T} A x|^{2},
 \]
where $x^{T}$ is the transpose of $x$ (without conjugation).

By singular value decomposition (see~\cite{Bhatia96}), every matrix can be
written in the form $V D U$ where $D$ is a nonnegative diagonal matrix and
$U, V$ are unitary transformations.
If the elements of $D$ are all distinct, positive and in decreasing order
then $U, V$ are unique.
In this case, clearly if $A$ is symmetric then $V=U^{T}$.
This can be generalized to the case when the elements of $D$ are not all
positive and distinct, using for example limits.
Thus, $A = U^{T} D U$.
This gives $x^{T} A x = x^{T} U^{T} D U x = (Ux)^{T} D (Ux)$.
As $x$ runs through all possible vectors with $\sum_{i} |x_{i}|^{2} =1$, so
does $Ux$.
Let $d_{1}$ be the largest element on the diagonal of $D$, then
$d_{1}^{2} = \nm{A^{\dgr}A}$.
 \[
   |(Ux)^{T} D Ux| = |\sum_{i} d_{i} (Ux)_{i}^{2}|
\le \sum_{i} d_{i} |(Ux)_{i}|^{2} \le d_{1}
 \]
since $\sum_{i} |(Ux)_{i}|^{2} = 1$.
The maximum of $|(Ux)^{T} D (Ux)|^{2}$ is achieved by the element 
$x = U^{-1} \ket{\bg_{1}}$, and then it is $d_{1}^{2} = u(A)$.
 \end{proof}

  \begin{lemma}\label{l.algeb-ub}
For $0 < d < N'$, there is a computable semi-density matrix $\rho$ with
 \[
     \sup_{\ket{\psi} = \ket{\phi}\ket{\phi}}
 - \log \bra{\psi}\rho \ket{\psi} \le \log (N' - d) - \log (1 - u(d, N)).
 \]
 \end{lemma}
 \begin{proof}
Using the notation of Lemma~\ref{l.uA},
let $F$ be the subspace of dimension $d$ of vectors $\ag$
on which the minimum $u(d, N)$ is achieved.
Witn $P = 1 - F$, let $\rho$ be the semi-density matrix defined in the
proof of Theorem~\ref{t.in-small-set}.
Similarly to~\eqref{e.Hl-in-small-set} we have, for
any $\psi = \ket{\phi}\ket{\phi}$: 
 \[
  -\log \bra{\psi}\rho \ket{\psi}
    \le \log (N' - d) - \log (1 - \bra{\psi} F \ket{\psi}).
 \]
Note that $\bra{\psi} F \ket{\psi} = |\braket{\ag}{\psi}|^{2}$ for some
$\ag \in F$, hence 
by~\eqref{e.uA} we have $\bra{\psi} F \ket{\psi} \le u$, hence the last
term of the right-hand side is $\le -\log (1 - u)$.
 \end{proof}

 \begin{proof}[Proof of Theorem \protect\ref{t.algeb}]
The reasoning of Theorem~\ref{t.cloning} implies that $\log N'$
lower-bounds the left-hand side in the above lemma.
Thus,
 \[
  \begin{split}
   \log N' & \le \log (1 - d/N') + \log N' - \log (1 - u),
\\      u  & \ge d / N'.
  \end{split}
 \]
 \end{proof}

\section{Randomness tests}

\subsection{Universal tests}

In classical algorithmic information theory (see for
example~\cite{LiViBook93}),
description complexity helps clarify what experimental outcomes
should be called random with respect to a hypothetical probability
distribution.
If the set of possible outcomes is a discrete one, say 
the set of natural numbers, then,
given a probability distribution $\nu$, we call a lower semicomputable
function $f(x)$ a \df{randomness test} if $\sum_{x} f(x) \nu(x) \le 1$.
It is known that there is a universal test $t_{\nu}(x)$, a test that
dominates all other tests to within a multiplicative constant.
An outcome is considered non-random with respect to $\nu$ 
when $t_{\nu}(x)$ is large.
In case of a computable distribution $\nu$, we have
 \begin{equation}\label{e.univ-test-expr}
   t_{\nu}(x) \eqm \frac{\m(x)}{\nu(x)},
 \end{equation}
where the multiplicative constant in the $\eqm$ depends on $\nu$.
(The general case is more complicated.)
The deficiency of randomness is defined as $d_{\nu}(x) = \log t_{\nu}(x)$.
In case of a computable distribution $\nu$ it is known to be
 \begin{equation}\label{e.def-rand-expr}
   \eqa - \log \nu(x) + \log \m(x) \eqa -\log \nu(x) - K(x)
 \end{equation}
Thus, for a computable distribution, 
the universal test measures the difference betwen the logarithm of
the probability and the complexity.

In the quantum setting, 
what corresponds to a probability distribution is a computable density
matrix $\rho$.
What corresponds to a function is a self-adjoint operator.
So, let us say that a \df{randomness test} is a lower semicomputable
self-adjoint operator $F_{\rho}$ with
 \[
   \tr F_{\rho} \rho \le 1.
 \]

 \begin{remark}\label{r.infinity}
In the theorem below, the expression 
 \[
  T' = \rho^{-1/2} \bmu \rho^{-1/2} 
 \]
appears, which does not make sense if $\rho$ is not invertible.
However, let us write $\sg = \mu^{1/2}\rho^{-1/2}$; this expression
makes sense on the subspace $V$ orthogonal to the kernel of $\rho$, and
therefore $T' = \sg^{\dgr}\sg$ also makes sense there.
Therefore we define $\bra{\psi} T' \ket{\psi}$ as $\infty$ for any $\ket{\psi} \notin V$,
and there is no problem for $\ket{\psi} \in V$.
 \end{remark}

 \begin{theorem}[Universal test]\label{t.univ-test}
There is a test $T_{\rho}$ which is \df{universal}
in the sense that it dominates each other test $R$: we have
$R \lem T_{\rho}$,
where the multiplicative constant in $\lem$ may depend on $R$ and $\rho$.
We have $T_{\rho} \eqm T'_{\rho} \eqm T''_{\rho}$ where
 \begin{align*}
      T'_{\rho} &= 
    \sum_{\ket{\phi}} \frac{\m(\ket{\phi}) \ket{\phi}\bra{\phi}}
                      {\bra{\phi} \rho \ket{\phi}},
\\    T''_{\rho} &= \rho^{-1/2} \bmu \rho^{-1/2}.
 \end{align*}
 \end{theorem}
 \begin{proof}
The proof of the existence of a universal test is similar to the proof of
Proposition~\ref{p.univ}.
The proof of $T \eqm T'$ is similar to the one showing $\bmu' \eqm \bmu$
in Theorem~\ref{t.univ-dens}.

Let us prove $T \eqm T''$.
To see that $T''$ is lower semicomputable, note that as direct computation
shows, for any operator $C$ the 
function $A \mapsto C^{\dgr} A C$ is monotonic on
the set of self-adjoint operators $A$ with respect to the relation $\le$.
By the cyclic property of the trace, we also have 
$\tr T'' \rho = \tr \bmu \le 1$.
This proves $T'' \lea T$, it remains to prove that $T \lem T''$.
This is equivalent to
 \[
  \rho^{1/2}T\rho^{1/2} \le \rho^{1/2}T''\rho^{1/2} = \bmu.
 \]
But the left-hand side is a
lower semicomputable nonnegative definite matrix whose trace is $\le 1$, 
again due to the cyclic property of trace.
Therefore by the defining property of $\bmu$, it is $\lem \bmu$.
 \end{proof}

The expression for $T''_{\rho}$ is similar to
~\eqref{e.univ-test-expr}, but it does not
separate the roles of the density
matrix $\rho$ and of the universal probability $\bmu$ as neatly, certainly 
not in the typical cases when $\bmu$ and $\rho$ do not commute.
Assume that the eigenvalues of $\rho$ are $p_{1} \ge  p_{2} \ge \dotsb$,
with the corresponding eigenvectors $\ket{v_{i}}$
(these exist since our space is finite-dimensional).
Let $(m_{ij})$ be the matrix of the operator $\bmu$ when expressed in this 
basis.
For a certain state $\ket{\psi} = \sum_{i} c_{i} \ket{v_{i}}$,  
we can express the value of the test on $\ket{\psi}$ as follows.
If there is any $i$ with
$p_{i}=0$ and $c_{i} \ne 0$ then according to Remark
~\ref{r.infinity}, the value is $\infty$.
Otherwise, it is
 \begin{equation}\label{e.test-expr-detail}
 \bra{\psi} T''_{\rho} \ket{\psi} 
  = \sum_{i,j} m_{ij}(p_{i}p_{j})^{-1/2}c_{i}^{*}c_{j}.
 \end{equation}
The term $(p_{i}p_{j})^{-1/2}c_{i}^{*}c_{j}$ is defined to be
0 if $c^{*}_{i}c_{j} =0$, and we excluded the case when
$p_{i}p_{j} = 0$ but $c^{*}_{i}c_{j} \ne 0$.
The roles of $\bmu$ and $\rho$ do not seem to be separable in the same way
as in the classical case.
However, if
$\rho$ is the uniform distribution then the expression simplifies to 
 \[
  N^{-1} \sum_{i,j=1}^{N} m_{ij} c_{i}^{*}c_{j} 
  = N^{-1} \bra{\psi}\bmu\ket{\psi},
 \]
which is the classical comparison of the probability
to the universal probability.

\subsection{Relation to Martin-L\"of tests}

The sum for $T'_{\rho}$ in Theorem~\ref{t.univ-test}
is similar to $\bmu'$ in Theorem~\ref{t.univ-dens}.
In the classical case and with a computable $\rho$, just like there, it can
be replaced with a supremum.
In the quantum case it cannot:
indeed, the expression of $\bmu'$ is a special case of $T'$, and we
have shown in Section~\ref{s.complexity-properties} that the sum in
$\bmu'$ cannot be replaced with supremum.
We do not know whether there is still an approximate relation like
in Theorem~\ref{t.Kq-Hl}: the proof does not carry over.

It is worth generalizing the sum for $T'_{\rho}$ as 
 \[
       \sum_{F} \frac{\m(F) F}{\tr F \rho}
 \]
where $F$ runs through all elementary nonnegative self-adjoint operators.
An interesting kind of self-adjoint operator is a projection $P$ to some
subspace.
Such a term looks like
 \[
   \frac{\m(P)}{\tr P \rho} P.
 \]
This term is analogous to a Martin-L\"of test.
An outcome $x$ would be caught by a Martin-L\"of test in the discrete
classical case if it falls into some simple set $S$ with small probability.
The fact that $S$ is simple means that $K(S)$ is small, in other words
$\m(S)$ is large.
Altogether, we can say that $x$ is caught if the expression
 \[
   \frac{\m(S)}{\rho(S)}1_{S}(x)
 \]
is large, where $1_{S}(x)$ is the indicator function of the set $S$.
In the quantum case, for state $\ket{\psi}$, what corresponds to this is
the expression
 \[
   \frac{\m(P)}{\tr P \rho} \bra{\psi}P\ket{\psi}.
 \]
The probability of $S$ translates to $\tr P \rho$, and $1_{S}(x)$
translates to $\bra{\psi}P\ket{\psi}$.
Thus, a quantum Martin-L\"of test catches a state $\ket{\psi}$ if it is
``not sufficiently orthogonal'' to some simple low-probability subspace.
Compare this with Theorem~\ref{t.in-small-set}.

As we see, the universal quantum randomness test contains the natural
generalizations of the classical randomness tests, but on account of the
possible non-commutativity between $\rho$ and $\bmu$, it may also test
$\ket{\psi}$ in some new ways that do not correspond to anything classical.
It would be interesting to find what these ways are.

\section{Proof of Theorem \protect\ref{t.Kq-lb}}\label{s.Kq-proof}

Let us denote 
 \[
  K_{m}(\ket{\psi}) = \min \setof{l(p) : U(p) = \ket{\phi},\;
      - \log |\braket{\phi}{\psi}|^{2} \le m}.
 \]
The first lemma lowerbounds $K_{\infty}(\ket{\psi})$, the later ones lowerbound
$K_{m}(\ket{\psi})$ for finite $m$.

 \begin{lemma}\label{l.K-infty}
For each $k$ there is a subspace $V$ of $\cQ_{n}$, of dimension
$2^{n} - 2^{k}$ with the property that for all $\ket{\psi} \in V$ we have 
$K_{\infty}(\ket{\psi}) \ge k$.
 \end{lemma}
 \begin{proof}
Let $p_{1}, \dots, p_{r}$ be all programs of length $< k$ for which 
$U(p_{m}) \in \cQ_{n}$.  Then $r < 2^{k}$.
Let $V$ be the set of elements of $\cQ_{n}$ orthogonal to all vectors 
of the form $U(p_{i})$.
 \end{proof}

Let $b_{n}$ denote the volume of the unit ball in an $n$-dimensional
Euclidean space.
Then for the surface volume $s_{n}$ of this ball we have
 \begin{equation}\label{e.surf-vol}
  b_{n-1} < s_{n} = n b_{n}.
 \end{equation}
For an angle $\ag$, let $s_{n}(\ag)$ be the surface volume of a subset of
the surface cut out by a cone of half-angle $\ag$:
for some vector $\ket{u}$, this is 
the set of all vectors $\ket{x}$ of unit length
with $\braket{u}{x} \ge \cos \ag$.
Thus, we have $s_{n} = s_{n}(\pi)$.
We are interested in how fast $s_{n}(\ag)$ decreases from $s_{n}/2$ to 0
as $\ag$ moves from $\pi/2$ to 0.
 
 \begin{lemma}
Let $\ag = \pi/2 - y$.
Then 
 \begin{equation}\label{e.cone}
 s_{n}(\ag) / s_{n} \lem \exp(-n y^{2}/2 + \ln n).
 \end{equation}
 \end{lemma}
 \begin{proof}
We have, for $k \ge 2$:
 \begin{equation}\label{e.volume-formula}
 s_{k}(\ag) = s_{k-1} \int_{0}^{\ag} \sin^{k-2} x\, d x
            \le s_{k-1} \ag \sin^{k-2} \ag.
 \end{equation}
So, we need to estimate $\int_{0}^{\ag} \sin^{n} x\, dx$.
The method used (also called ``Laplace's'' method),
works for any twice differentiable function with a single maximum.
Let $g(x) = \ln \sin x$, then it can be checked that $g'(\pi/2) = 0$,
$g''(\pi/2) = -1$, $g'''(x) > 0$ for $x < \pi/2$.
The Taylor expansion around $\pi/2$ gives, for $y > 0$:
 \[
   g(\pi/2 - y) = -y^{2}/2 - y^{3} g'''(\pi/2 - z)/6 < -y^{2}/2.
 \]
where $0 < z < y$.
Hence, since $\sin x$ is increasing, we have for $x < \pi/2 - y$,
 \[
   \sin^{n} (x) < e^{-n y^{2}/2}.
 \]
On the other hand, by~\eqref{e.surf-vol}, $s_{k} \ge b_{k-1} = s_{k-1}/(k-1)$,
showing $s_{n-1} < (n-1) s_{n}$.  Hence
 \[
   s_{n}(\ag) < \frac{\pi}{2}e^{-(n-2) y^{2}/2} s_{n-1} 
              < \frac{(n-1)\pi}{2}e^{-(n-2) y^{2}/2} s_{n}
              \lem  s_{n} e^{-n y^{2}/2 + \ln n}.
 \]
 \end{proof}
 
 \begin{lemma}\label{l.K-finite}
In any Hilbert space $\cH$ of dimension $2^{n}$ (it may be a subspace of
some $\cQ_{r}$), the
volume fraction of the set of unit vectors
$\ket{\psi}$ in $\cH$ with the property that $K_{m}(\ket{\psi}) < k$ is
 \[
   \lem \exp(-2^{n - m} + k\ln 2 + n).
 \]
 \end{lemma}
 \begin{proof}
We view $\cQ_{n}$ as a $2^{n+1}$-dimensional Euclidean space.
Assume $-\log |\braket{\phi}{\psi}|^{2} \le m$.
If $\ag$ is the angle between $\ket{\phi}$ and $\ket{\psi}$ then this means
 \[
  2^{-m/2} < |\braket{\phi}{\psi}| = \cos \ag  = \sin(\pi / 2 - \ag)
  \le \pi/2 - \ag,
 \]
giving $\ag < \pi/2 - 2^{-m/2}$.
For a fixed $\ket{\phi}$, the relative volume (with respect to $s_{2^{n+1}}$)
of the set of vectors with $-\log |\braket{\phi}{\psi}|^{2} \le m$ 
is therefore by~\eqref{e.cone}
 \[
   \lem \exp(-2^{n+1} 2^{-m}/2 + n) = \exp(-2^{n - m} + n).
 \]
Let $p_{1}, \dots, p_{r}$ be all programs of length $< k$ for which 
$U(p_{m}) \in \cQ_{n}$.  Then $r < 2^{k}$.
The volume of all vectors $\ket{\psi}$ that are close in the above sense to at
least one of the vectors $U(p_{i})$ is thus 
 \[
   \lem 2^{k} \exp(-2^{n - m} + n ) 
   = \exp(-2^{n - m} + k\ln 2 + n).
 \]
 \end{proof}

 \begin{proof}[Proof of Theorem \protect\ref{t.Kq-lb}]
According to Lemma~\ref{l.K-infty}, there is a subspace $V$
of $\cQ_{n}$, of
dimension $2^{n} - 2^{n-1} = 2^{n-1}$ with the property that for all
$\ket{\psi} \in V$, for all $m$ we have $K_{m}(\ket{\psi}) \ge n - 1$.
Let $m = n - 2 \log n$.
We can apply Lemma~\ref{l.K-finite} to this subspace $V$ of dimension
$2^{n-1}$, and obtain that for a certain constant $c$,
the volume fraction of vectors with 
$K_{m}(\ket{\psi}) < 2 n$ is 
 \begin{align*}
   &\le \exp(-2^{(n - 1) - (n - 2 \log n)} + 2 n\ln 2 + (n - 1) + c)
\\ &= \exp(-n^{2}/2 + n(2\ln 2 + 1) + c - 1).
 \end{align*}
  If $n$ is large this is smaller than 1, so there are states $\ket{\psi}$ with 
$K_{\infty}(\ket{\psi}) \ge n - 1$ and $K_{n - 2\log n}(\ket{\psi}) > 2 n$.
For these, clearly 
 \[
   \Kq(\ket{\psi}) \ge (n - 1) + (n - 2\log n + 1) = 2 n - 2\log n.
 \]
 \end{proof}

\section{Conclusions}

We advanced a new proposal to extend the theory of
descriptional complexity to the quantum setting.
The approach starting from the universal density matrix appears to be
fruitful and leads to some attractive relations.
However, the theory is still very incomplete.
The following tasks seem to be the most urgent.

 \begin{enumerate}
  \item Strengthen Theorem~\ref{t.QClessH} in a way that the smallness of
$\Hu(\ket{\psi})$ allows a direct inference on the smallness of
$QC(\ket{\psi})$ (or find a counterexample).
For this, it seems to us that behavior of a monotonically increasing
sequence of density functions needs to be understood better: namely,
whether some approximate monotonicity can be stated about 
the subspaces $E_{k}$.
Even if such a monotonicity will be found, even if Thoerem~\ref{t.QClessH}
can be proved for $\bmu$ instead of just computable density matrices, the
result is too weak.
To strengthen it, probably the theory of
indeterminate-length quantum codes (the quantum analog of variable-length
codes) will be needed, as developed in~\cite{SchumWestmIndetLen00}.

  \item Find the proper generalization to the quantum setting
of the classical theorem saying that information cannot increase under the
effect of any probabilistic computable transformation.

  \item What kind of addition theorems can be expected for quantum
description complexity?
The question is unsolved even for the von Neumann entropy.
Also, the translation between
the results on quantum description complexity and those
on the von Neumann entropy will not be straightforward.
As we remarked, the relation
$\Hu(\ket{\phi}\ket{\psi}) \gea \Hu(\ket{\phi})$ holds
while $S(\rho_{X}) \le S(\rho_{XY})$ does not.
Still, maybe the study of the problem for quantum description complexity
helps with the understanding of the problem for von Neumann entropy, and
its relation to coding tasks of quantum information theory.

Despite all the caveats, let us ask the question (risking that somebody
finds a trivial answer): does $\Hu$ obey strong superadditivity?
 \end{enumerate}

\subsection*{Acknowledgement}

The author is grateful to Paul Vit\'anyi, Harry
Buhrman and Ronald de Wolf for discussions.


\end{document}